# Natural modes of the two-fluid model of two-phase flow

Alejandro Clausse[1,*] and Martín López de Bertodano[2,#]

[1]CNEA-CONICET and National University of Central Buenos Aires, 7000 Tandil, Argentina.

[2]School of Nuclear Engineering, Purdue University, West Lafayette, Indiana 47907,

**Abstract**

A physically-based method to derive well-posed instances of the two-fluid transport equations for two-phase flow, from the Hamilton principle, is presented. The state of the two-fluid flow is represented by the superficial velocity and the drift-flux, instead of the average velocities of each fluid. This generates the conservation equations of the two principal motion modes naturally: the global center-of-mass flow and the relative velocity between fluids. Well-posed equations can be obtained by modelling the storage of kinetic energy in fluctuations structures induced by the interaction between fluids, like wakes and vortexes. In this way, the equations can be regularized without losing in the process the instabilities responsible for flow-patterns formation and transition. A specific case of vertical air-water flow is analyzed showing the capability of the present model to predict the formation of the slug flow regime as trains of non-linear waves.

Two-phase flows are important transport processes both in natural and artificial domains. Examples of natural systems are the geysers, hot jets of water and steam ejected from volcanic soils. In industry, the uses of two-fluid flows are numerous, ranging from transport of petroleum mixtures in ducts, the variety of boiling heat transfer phenomena involved in energy systems, management of interfacial reactions in chemical processes, etc.

Multiple fluid mixtures are generally too complex to describe by keeping track of all interfaces and motions at both sides. We must resort then to representations of the system based on mean-field state variables and their correspondent evolution equations. The overall averaging procedure based on continuum mechanics, established long ago, provides with a consistent theoretical framework [1, 2]. However, its major theoretical development, the two-fluid model, where the phases are described as interwoven continuum fluids, yields an ill-posed mathematical problem, which, as a recent update states ,"despite extensive efforts to remedy, and despite claims to the contrary, remains to this day" [3]. Recently, strong evidence was presented indicating that the physical mechanism responsible for the problem is the Kelvin-Helmholtz instability (KHI), summoned whenever two fluids are forced to flow together shearing the pressure field [4]. This is another manifestation of KHI, one of the most important hydrodynamic instabilities, which continues drawing the attention of the fluid community [5, 6]. Current numerical regularization schemes basically suppress this instability. Alas, the price paid is that any possible predictive capacity of the physical phenomena associated to the instability, like flow pattern formation, is lost in the maneuver.

The motivation of our work is the solution of the ill-posedness problem of the two-fluid model by means of inertia-exchange mechanisms. We present a physically-based way to derive well-posed instances of the two-fluid transport equations from the first principles of field theory, which opens a promising road to focus future research efforts, especially for conflating experimental data and theoretical models. In the 80s and 90's several researchers applied the Hamilton principle of mechanics to the one-dimensional dispersion of one fluid into the continuous bulk of another. A variety of inertial coupling models were explored and different lines of interpretation of the resulting conservation equations were investigated [7-10].

Since most applications of two-fluid flows take place in long conduits, like pipes, ducts and channels, where mixtures are being transported, processed or heated, one-dimensional approaches are the most common for treating practical problems because they highly reduce computational costs. But there are also other factors, e.g., that less phenomenological parameters are required to close the equations and that most of them were developed for one-dimensional conditions.

We adopt the usual notation where the physical properties of each fluid are identified with the sub-index $i = 1, 2$. By convention, $i = 1$ is assigned to the denser fluid. All properties are assumed constant and, therefore, the fluids are incompressible. The volume fraction occupied by fluid-$i$ is treated as a state field of the system, and it is denoted by $\alpha_i$. The streamwise velocity of each fluid averaged over the volume occupied by the corresponding fluid is denoted by $u_i$. Mass densities are denoted by $\rho$. Then, the following variables can be constructed with $\alpha_i$ and $u_i$:

$$\rho_m = \sum_i \alpha_i \rho_i \qquad \text{Mixture density} \qquad (1)$$

$$j = \sum_i \alpha_i u_i \qquad \text{Superficial velocity} \qquad (2)$$

$$\rho_m v_m = \sum_i \rho_i \alpha_i u_i \qquad \text{Mixture momentum} \qquad (3)$$

$$u_r = u_2 - u_1 \qquad \text{Relative velocity} \qquad (4)$$

$$J = u_r \alpha_1 \alpha_2 \qquad \text{Drift flux} \qquad (5)$$

The superficial velocity, $j$, is the velocity of the center of volume of the mixture, whereas the velocity $v_m$ is the velocity of the center of mass of the mixture. The drift flux, $J$, is the volumetric flux of fluid-2 passing through a cross-section surface that moves with velocity $j$.

The continuity equation of each fluid is

$$\frac{\partial \alpha_i}{\partial t} + \frac{\partial \alpha_i u_i}{\partial x} = 0 \tag{6}$$

which can be combined in the two following forms:

$$\frac{\partial \alpha_i}{\partial t} + j\frac{\partial \alpha_i}{\partial x} + (-1)^i \frac{\partial J}{\partial x} = 0 \tag{7}$$

$$\frac{\partial j}{\partial x} = 0 \tag{8}$$

The kinetic energy density of a representative volume element (RVE) of a two-fluid system is:

$$K = \frac{1}{2}\sum_i \rho_i \alpha_i u_i^2 + \frac{1}{2}\sum_i \rho_i \alpha_i \sigma_i^2 \tag{9}$$

where:

$$\sigma_i^2 = \langle u_i^2 \rangle - u_i^2 \tag{10}$$

is the mean quadratic deviation of the velocity of fluid-$i$, and $\langle u_i^2 \rangle$ is the square of the velocity module of each fluid averaged over the volume occupied by the corresponding fluid.

At this level of description, the state of the system is represented by the fields $u_i$ and $\alpha_i$. However, we are interested in describing the two-fluid flow with variables representing the global motion of the mixture and the relative motion between fluids. To do so, we express the Lagrangian in terms of the superficial velocity, $j$, and the drift flux, $J$. Eq. (9) then becomes:

$$K = \frac{1}{2}\rho_m j^2 + \frac{1}{2}\Gamma J^2 - \Delta\rho\, jJ \tag{11}$$

where:

$$\Gamma = \sum_i \frac{\rho_i}{\alpha_i} + \sum_i \rho_i \alpha_i \gamma_i(\alpha_i) \tag{12}$$

$$\gamma_i(\alpha_i) = \frac{1}{\alpha_i^2(1-\alpha_i)^2}\left(\frac{\sigma_i^2}{u_r^2}\right) \tag{13}$$

In principle, $\sigma_i^2$ may have a turbulent component proportional to $u_i^2$, but this would be dissipative in nature, and so it will be incorporated later as a dissipative force. The part of $\sigma_i^2$ that interests us is related to the velocity redistribution due to the interaction between fluids, and in general is proportional $u_r^2$ with a coefficient that depends on the volume fraction $\alpha_i$ of the correspondent fluid [7, 9]. This term is analogous to the added mass of immersed bodies. The Lagrangian of the two-fluid system is then written as:

$$L = K + \sum_n \varphi_n R_n \tag{14}$$

where $R_n$ are admissibility conditions that restrain the variation of the state variables, $\alpha_i$, $j$ and $J$; and $\varphi_n$ are the corresponding Lagrange multipliers. In our case, the restrictions are the continuity Eqs. (7) and the confinement relation:

$$\sum_i \alpha_i = 1 \tag{15}$$

Then:

$$L = \frac{1}{2}\rho_m j^2 + \frac{1}{2}\Gamma J^2 - \Delta\rho\, jJ + \sum_i \varphi_i \left[\frac{\partial \alpha_i}{\partial t} + j\frac{\partial \alpha_i}{\partial x} + (-1)^i \frac{\partial J}{\partial x}\right] + \varphi_\alpha \sum_i \alpha_i \tag{16}$$

The corresponding Euler-Lagrange equations may now be cast in two modes [17]. For the center of mass of the mixture:

$$\frac{\partial \mathcal{M}_m}{\partial t} + \frac{\partial \Sigma_m}{\partial x} = 0 \tag{17}$$

where:

$$\mathcal{M}_m = \rho_m v_m \tag{18}$$

$$\Sigma_m = \rho_m v_m^2 + \frac{\rho_1 \rho_2}{\alpha_1 \alpha_2 \rho_m} J^2 + \frac{1}{2} J^2 \sum_i \rho_i \frac{d\gamma_i}{d(1/\alpha_i)} + p \tag{19}$$

The first term of $\Sigma_m$ is the global momentum flux of the mixture, which behaves as a pseudo-fluid with density $\rho_m$ and velocity $v_m$. The second term is the momentum flux due to the relative motion without inertial coupling between fluids. The third term is the momentum flux introduced by the inertial coupling between fluids. The last term of $\Sigma_m$ is the pressure, which is recognized as the scalar field $p = -\varphi_\alpha$. If $\gamma_i = 0$, Ishii's drift-flux mixture momentum balance equation is recovered [1].

For the relative motion between fluids, we have:

$$\frac{\partial \mathcal{M}_J}{\partial t} + \frac{\partial \Sigma_J}{\partial x} = 0 \tag{20}$$

where:

$$\mathcal{M}_J = \Gamma J - j\Delta\rho \tag{21}$$

$$\Sigma_J = -\frac{1}{2}\Gamma' J^2 + \Gamma jJ \tag{22}$$

Henceforth the prime indicates derivative respect to $\alpha_2$, and $\Delta\rho = \rho_1 - \rho_2$. The length scale of this equation is of the order of the pipe diameter, and the characteristic time is much shorter than the transit time along the conduit. Eqs. (20)-(22) have some interesting associations. By imposing $j = 0$, Eqs. (20)-(22) coincide with the equation found in previous works for that condition [9, 10]. Furthermore, if $j = 0$ and $\rho_2 = 0$, then $\mathcal{M}_J =$

$\rho_1 J/\alpha_1$ and $\Sigma_J = (\rho_1/2)(J/\alpha_1)^2$, and Eq. (20) reduces to the Burger's equation. Also, in stratified flow, a gravitational term appears, which for $j = 0$ and $\rho_2 = 0$ leads to the shallow water equation of Saint-Venant.

Perturbing and linearizing Eqs. (20)-(22) yields the following dispersion relation:

$$\frac{\omega}{k} = j_o + J_o \Gamma_o \left( -Q'_o \pm \sqrt{\frac{1}{2} Q_o Q''_o} \right) \tag{23}$$

Where, the sub-index $o$ refers to the steady-state and $Q_o = 1/\Gamma_o$. Since $Q_o$ is always positive, if $Q''_o < 0$ the system is elliptic for any wave number. In particular, for $k = \infty$ (*i.e.*, zero wave length) the perturbation growth rate is infinite. This sort of ultraviolet divergence is the infamous ill-posedness problem, which makes it impossible to achieve numerical convergence. The necessary condition of a well-posed two-fluid theory reduces to:

$$Q''_o \geq 0 \tag{24}$$

or equivalently:

$$2\Gamma'^2_o - \Gamma_o \Gamma''_o \geq 0 \tag{25}$$

Eq. (25) is the same condition found for constant $j$ [9]. Our result shows that the criterion given by Eq. (24) is general for the incompressible two-fluid model, holding also if $j$ varies in time. This is a significant result, for it proves that the cause of the ill-posedness problem lies in the relative fluid motion mode of the system, particularly in the term $\frac{1}{2}\Gamma J^2$ of the Lagrangian. Furthermore, Eq. (25) is equivalent to the threshold of the Kelvin-Helmholtz instability in two-layers shallow water [11], appearing here as an extreme manifestation that triggers the growth of any wave length, thus rendering the numerical solution dependent on the spatial resolution.

It should be mentioned that in horizontal flows, gravity and surface tension can make the system hyperbolic up to certain values of $J$ while stratified conditions hold [12, 13]. Those special conditions are not difficult to introduce into Eq. (23) to extend the criterion. Nevertheless, in most practical cases, the role played by the magnitude $Q$ appears to be the crux of the matter for a well-posed model, in the sense that numerical convergence to the differential limit be achievable. However, such well-posed models can still be unstable due to the interaction between inertia and drag. This is a mechanism which is well known in the global motion of the mixture, especially when boiling is included: the so-called density-wave instabilities. In the relative-motion mode between fluids, the dynamic interplay between the interfacial drag and the inertia coupling can also lead to destabilizing well-posed mechanisms.

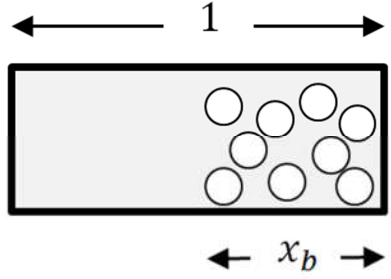

Figure 1. Representative volume element used in the model of inertial coupling. The bubble cluster occupies a mean fraction $\alpha_b$ of the channel cross-section.

Next we will present a simple model of inertial coupling to show the procedure. Consider the RVE depicted in Fig. 1. A bubbly cluster or a wide bubble of the lighter fluid, occupies a fraction $\alpha_b$ of the channel cross section along a length fraction $x_b$ of the RVE. The remaining portion of the RVE is a slug filled with the heavier fluid moving streamwise with mean velocity $u_s$. Note that the volume fraction of fluid-2 is:

$$\alpha_2 = \alpha_b x_b \tag{26}$$

Disregarding fluctuations having zero-average streamwise velocity, like vortexes, the kinetic energy density in the RVE is given by:

$$K = \frac{1}{2}\rho_1(1-x_b)u_s^2 + \frac{1}{2}\rho_1 x_b(1-\alpha_b)u_f^2 + \frac{1}{2}\rho_2\alpha_2 u_2^2 \tag{27}$$

where $u_f$ is the average velocity of fluid-1 in sector $x_b$. Using Eqs. (9)-(13), we find that the inertial coupling is given by [8, 10]:

$$\gamma_1 = \frac{\alpha_b - 1 + \alpha_1}{\alpha_1^2(1-\alpha_1)(1-\alpha_b)} \tag{28}$$

Now suppose that there are zero-average velocity fluctuations in fluid-1, like undulating coherent structures at the back and amid the bubble cluster. Although these fluctuations do not affect $u_1$, they do contribute to $\langle u_1^2 \rangle$. In certain ways, this would be similar to the thermodynamic degradation of mechanical energy into thermal energy, although in this case the pseudo-thermalization takes place between the scale of the tube length and the scale of the tube diameter. A simple model for this mechanism is to assume that the energy of these wavy fluctuations is proportional the kinetic energy of the streamwise velocity dispersion in the RVE, represented by Eq. (28), and to the room available for the development of these perturbations, i.e., $\alpha_1$. If the coupling coefficient of proportionality is a constant parameter $w$, then:

$$\gamma_1 = \frac{(1+w\alpha_1)(\alpha_b - 1 + \alpha_1)}{\alpha_1^2(1-\alpha_1)(1-\alpha_b)} \tag{29}$$

The inertial coupling is then characterized by the control parameters $w$ and $\alpha_b$. The parameter $w$ represents the ratio of the kinetic energy of zero-average fluctuations of fluid-

1 divided by the quadratic deviation of its streamwise velocity. The parameter $\alpha_b$ characterizes local axial asymmetries of the distribution of fluid-2 in the RVE.

We can verify that:

$$\Gamma'^2 - \frac{1}{2}\Gamma\Gamma'' = \frac{w\rho_1}{(1-\alpha_b)(1-\alpha_1)^3}\left[\rho_2 + \rho_1\left(\frac{\alpha_b}{1-\alpha_b}\right)(1+w)\right] \qquad (30)$$

which is always positive, rendering the system hyperbolic and thus well-posed. Higher order models of inertial coupling can be addressed by refining the generation mechanisms of coherent structures. It should be noted that, in spite of being hyperbolic, the system may still be linearly unstable, because of the interaction between drag and inertial forces.

To verify the performance of the derived modal two-fluid equations, we consider an air-water upward flow in a circular conduit of length $L = 4$ m. Constant superficial velocity $j$ is assumed, and therefore Eqs. (17) and (20) decouple. In this way, the study focuses on the inter-fluid instability, leaving aside the influence of the acceleration of the global mixture. On account of Eq. (8), this condition is easy to achieve experimentally by controlling the fluids injection at the inlet.

Sources of gravity and interfacial drag are added in the drift-flux momentum equation as:

$$\frac{\partial \Gamma J}{\partial t} + \frac{\partial \Sigma_J}{\partial x} = g\Delta\rho - \frac{C_D}{D}\rho_1 u_r |u_r| \qquad (31)$$

where $D = 30$ cm is the tube diameter, and:

$$C_D = \exp(1.32 + 8.4\alpha_2 - 42\,\alpha_2^2 + 45\alpha_2^3 - 15.4\alpha_2^4) \qquad (32)$$

is an empirical interfacial drag coefficient we produced from experimental data [17].

Eq. (31) together with the continuity Eq. (7) are discretized in space using a fourth order estimator of the spatial derivatives, and the resulting set of ordinary differential equations in time is solved by means of a backward differentiation scheme. For convenience, periodic boundary conditions are imposed to the spatial domain, which is an idealized representation of fully developed flows.

To better visualize the non-linear behavior of the system, it is useful to define the system variable:

$$\Delta\alpha_2 = \int_0^{L/2} \alpha_2 dx - \int_{L/2}^{L} \alpha_2 dx \qquad (33)$$

representing the imbalance of gas volume in the conduit; $\Delta\alpha_2 = 0$ indicating that the gas mass is sheared in exact halves between $x \in (0, L/2)$ and $x \in (L/2, L)$.

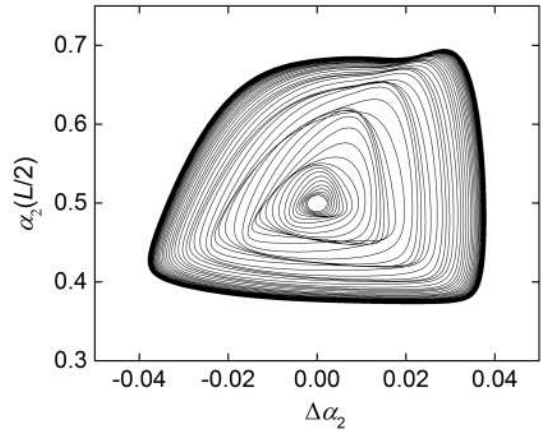

Figure 2. Phase-space evolution of the simultaneous values of $\Delta\alpha_2$ and $\alpha_2$ at $x = L/2$, for the reference case: $L = 4$ m, $D = 30$ cm, $j = 0.5$ m/s, $\langle\alpha_2\rangle = 0.5$, $\alpha_b = 0.9$, $w = 0.4$. The transient starts with a small perturbation around the steady-state values. The system develops an oscillation with growing amplitude, approaching asymptotically to a limit cycle.

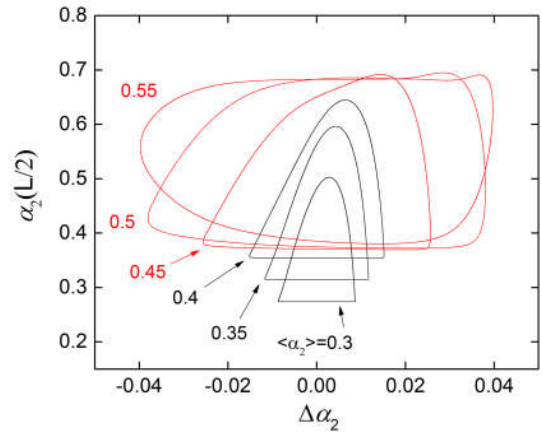

Figure 3. Limit cycles for the reference parameters ($\alpha_b = 0.9$, $w = 0.4$) and different background $\langle\alpha_2\rangle$. The cycles up to $\langle\alpha_2\rangle = 0.4$ are associated to bubble clustering, higher $\langle\alpha_2\rangle$ are associated to slug-flow pattern.

Figure 2 shows the phase-plane evolution of $\Delta\alpha_2$ and the simultaneous value of $\alpha_2$ at $x = L/2$, for the case with background $\langle\alpha_2\rangle = 0.5$. It can be seen that the initial small perturbation grows until non-linear mechanisms act containing the amplitude of the oscillation. The dynamics has the properties of a soliton, in the sense that the asymptotic state is independent of the initial perturbation. In the phase-plane, the asymptotic trajectory describes a limit cycle. Figs. 3 and 4 depict the limit cycles obtained for different background gas fractions $\langle\alpha_2\rangle$ between 0.3 and 0.55, and the corresponding snapshots of spatial profiles $\alpha_2(x)$. For $\langle\alpha_2\rangle$ between 0.3 and 0.4, the profiles present narrow spikes of gas concentrations, which can be construed as small clusters of bubbles. For $\langle\alpha_2\rangle$ between 0.45 and 0.55, the regions of higher gas fraction become longer, which can be construed as Taylor bubbles occupying the center of the conduit cross section surrounded by a thin film of liquid. These bubbles are separated by slugs of low gas concentration, typical of the slug flow-pattern regime. This result is in agreement with experimental observations [14, 15]. Furthermore, by increasing the resolution of the spatial discretization, 99% convergence was achieved with 200 nodes.

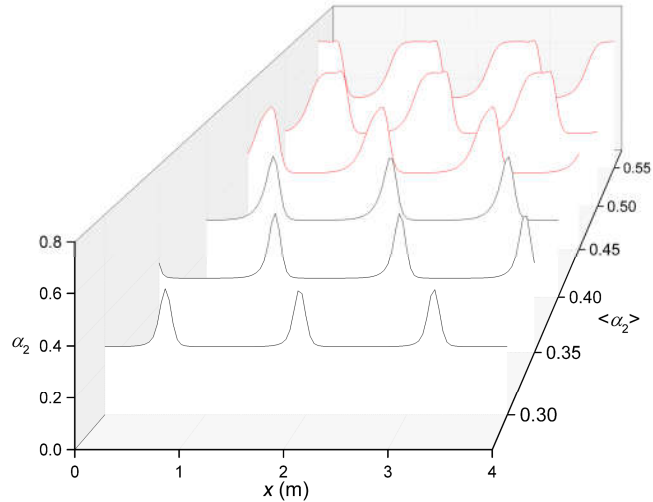

Figure 4. Gas fraction profiles for the reference parameters ($\alpha_b = 0.9$, $w = 0.4$) and different background $\langle\alpha_2\rangle$. Profiles up to $\langle\alpha_2\rangle = 0.4$ are associated to bubble clustering, higher $\langle\alpha_2\rangle$ are associated to slug-flow pattern.

In the resulting scenario, the control of the character of the equations (elliptic, hyperbolic, or parabolic) is achieved by the kinetic energy of the coherent (or pseudo-turbulent) structures of the fluid fluctuations, like wakes, vortexes, localized flow gradients, etc. This is performed via the inertial coupling function $\gamma_i(\alpha_i)$ given by Eq. (13). It is interesting to reckon that similar expressions of inertial coupling in mixtures of continuous media were found by using the principle of virtual powers [16]. Different spatial configurations of the fluids are associated to different inertial coupling functions, which in turn yield different energy partitions. In principle, the system will "chose" the configurations that minimize the action. Therefore, the new methodology regularizes the equations using a physically based procedure. Essentially, the method consists of producing parametric families of inertial coupling functions, providing enough flexibility for the system to find equilibrium configurations that match the experimental observations.

A brief comment about the physical interpretation of the inertial coupling function is offered as an educated conjecture. Notice that the actual effect of the inertia coupling can be construed as a modification of the apparent density of the fluids, as if a fraction of the mass of each fluid were exchanged in the process. Viewed from this perspective, regarding that ultimately the inertial terms translate into force terms in the momentum transport equations, the inertial coupling resembles in a certain way the physics of elementary particles, where the interactions take the form of force carriers. Suggestively, the present modal two-fluid field theory permits the construction of inertial exchange models in a progressive way, much as the way of the perturbative levels in which elementary force-carrier particles appear. Actually, note that the function $\Gamma$ can be (loosely) seen to play a role similar to the propagator in field theory. By way of illustration, Fig. 5 lists the inertial coupling function $\Gamma$ obtained at different levels of perturbation for the model given by Eqs. (12), (28) and (29). By the same token, other forms of perturbation of the velocity fields can be further introduced and tested against experimental observations, analogous to how the experiments in particle colliders provide insight to field theories. Among the mechanisms that are promising candidates for future studies are: bubble entrainment, collision, breakup and coalescence, interaction with solid structures, interfacial deformation and waves, vortexes inside or in the wake of bubbles, among others.

| Configuration | $\Gamma$ |
|---|---|
| 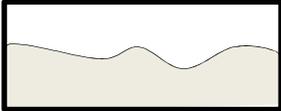 | $\Gamma_1 = \dfrac{\rho_1}{\alpha_1} + \dfrac{\rho_2}{\alpha_2}$ |
| 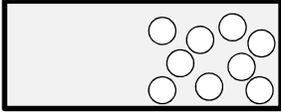 | $\Gamma_2 = \Gamma_1 + \dfrac{\rho_1}{\alpha_1 \alpha_2}\left(\dfrac{\alpha_s - \alpha_2}{1 - \alpha_s}\right)$ |
| 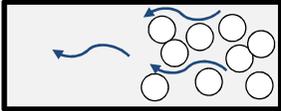 | $\Gamma_3 = \Gamma_2 + w\dfrac{\rho_1}{\alpha_2}\left(\dfrac{\alpha_s - \alpha_2}{1 - \alpha_s}\right)$ |

Figure 5. Inertial function $\Gamma$ in successive levels of detail of kinetic energy. The conservation equations are elliptic with $\Gamma_1$, parabolic with $\Gamma_2$, and hyperbolic with $\Gamma_3$.

**Acknowledgements**
The authors want to thank Alexander López de Bertodano for his interesting discussions. This work was partially supported by the Argentine National Atomic Energy Commission.

*clausse@exa.unicen.edu.ar

#bertodan@purdue.edu

## Supplementary Material for "Natural modes of the two-fluid model of two-phase flow"

### Derivation of the Euler-Lagrange equations of the modal two-fluid model

We start with the Lagrangian given by:

$$L = \frac{1}{2}\rho_m j^2 + \frac{1}{2}\Gamma J^2 - \Delta\rho\, jJ + \sum_i \varphi_i \left[\frac{\partial \alpha_i}{\partial t} + j\frac{\partial \alpha_i}{\partial x} + (-1)^i \frac{\partial J}{\partial x}\right] + \varphi_\alpha \sum_i \alpha_i \quad \text{(SM-1)}$$

which accounts for the kinetic energy in terms of $j$, $J$ and $\alpha_i$, and the admissibility conditions of those variables with the Lagrange multipliers $\varphi_i$ and $\varphi_\alpha$.

The following gauging is made to simplify the derivation:

$$L \to L - \sum_i \left[\frac{\partial \alpha_i \varphi_i}{\partial t} + j\frac{\partial \alpha_i \varphi_i}{\partial x} + (-1)^i \frac{\partial J \varphi_i}{\partial x}\right] \quad \text{(SM-2)}$$

so, the Lagrangian becomes:

$$L = \frac{1}{2}\rho_m j^2 + \frac{1}{2}\Gamma J^2 - \Delta\rho\, jJ - \sum_i \alpha_i \frac{\partial \varphi_i}{\partial t} - j\sum_i \alpha_i \frac{\partial \varphi_i}{\partial x} - J\sum_i (-1)^i \frac{\partial \varphi_i}{\partial x} \quad \text{(SM-3)}$$
$$+ \varphi_\alpha \sum_i \alpha_i$$

The corresponding Euler-Lagrange equations are:

$$\frac{\partial L}{\partial j} = \rho_m j - \Delta\rho\, J - \sum_i \alpha_i \frac{\partial \varphi_i}{\partial x} = 0 \quad \text{(SM-4)}$$

$$\frac{\partial L}{\partial J} = \Gamma J - \Delta\rho\, j - \sum_i (-1)^i \frac{\partial \varphi_i}{\partial x} = 0 \quad \text{(SM-5)}$$

$$\frac{\partial L}{\partial \alpha_i} = \frac{1}{2}\rho_i j^2 + \frac{1}{2}\Gamma_i J^2 - \frac{\partial \varphi_i}{\partial t} - j\frac{\partial \varphi_i}{\partial x} + \varphi_\alpha = 0 \quad \text{(SM-6)}$$

where $\Gamma_i \equiv \partial\Gamma/\partial\alpha_i$. Eqs. (SM-4)-(SM-6) form a linear algebraic system for $\partial\varphi_i/\partial x$ and $\partial\varphi_i/\partial t$, which can be cast in two compact $2\times 2$ sets as:

$$\mathbf{M}\,\phi = \beta \quad \text{(SM-7)}$$

where:

$$\mathbf{M} = \begin{pmatrix} \alpha_1 & \alpha_2 \\ -1 & 1 \end{pmatrix} \quad \text{(SM-8)}$$

and $\beta = (a,b)^T$ for $\phi = \left(\frac{\partial \varphi_1}{\partial t}, \frac{\partial \varphi_2}{\partial t}\right)^T$, $\beta = (c,d)^T$ for $\phi = \left(\frac{\partial \varphi_1}{\partial x}, \frac{\partial \varphi_2}{\partial x}\right)^T$; with:

$$a = -\frac{1}{2}\rho_m j^2 + jJ\Delta\rho + \frac{1}{2}\Gamma_m J^2 + \varphi_\alpha \tag{SM-9}$$

$$b = \frac{1}{2}\Delta\rho j^2 + \frac{1}{2}\Gamma' J^2 - \Gamma J j \tag{SM-10}$$

$$c = j\rho_m - J\Delta\rho = \rho_m v_m \tag{SM-11}$$

$$d = \Gamma J - \Delta\rho\, j \tag{SM-12}$$

and the following magnitudes are defined to simplify notation:

$$\Gamma' = \sum_i (-1)^i \Gamma_i \tag{SM-13}$$

$$\Gamma_m = \sum_i \alpha_i \Gamma_i \tag{SM-14}$$

The solutions are straightforward to obtain as:

$$\phi = \mathbf{M}^{-1}\beta = \begin{pmatrix} 1 & -\alpha_2 \\ 1 & \alpha_1 \end{pmatrix}\beta \tag{SM-15}$$

which finally gives:

$$\frac{\partial \varphi_i}{\partial t} = a - (-1)^i (1 - \alpha_i) b \tag{SM-16}$$

$$\frac{\partial \varphi_i}{\partial x} = c - (-1)^i (1 - \alpha_i) d \tag{SM-17}$$

Cross differentiating and subtracting Eqs. (SM-16) and (SM-17), yields to the master transport equation for each fluid:

$$\frac{\partial U_i}{\partial t} - \frac{\partial \Xi_i}{\partial x} = 0 \tag{SM-18}$$

where:

$$U_i = \rho_i\, j - [\Delta\rho - (-1)^i (1 - \alpha_i)\Gamma] J \tag{SM-19}$$

$$\Xi_i = jJ[\Delta\rho - (-1)^i (1 - \alpha_i)\Gamma] + \frac{1}{2}\Gamma_i J^2 + \varphi_\alpha \tag{SM-20}$$

**Equation of the momentum mixture**

Multiplying Eqs. (SM-18) by $\alpha_i$ and adding them up for both fluids, yields:

$$\sum_i \alpha_i \frac{\partial U_i}{\partial t} - \sum_i \alpha_i \frac{\partial \Xi_i}{\partial x} = 0 \tag{SM-21}$$

Using Eq. (SM-19), the first term of Eq. (SM-18) gives:

$$\sum_i \alpha_i \frac{\partial U_i}{\partial t} = \rho_m \frac{dj}{dt} - \Delta\rho \frac{\partial J}{\partial t} + \sum_i (-1)^i \alpha_i \frac{\partial}{\partial t}[(1-\alpha_i)\Gamma J] \quad \text{(SM-22)}$$

Using Eq.(3), Eq. (SM-22) can be written as:

$$\sum_i \alpha_i \frac{\partial U_i}{\partial t} = \frac{\partial \rho_m v_m}{\partial t} + j\frac{\partial \rho_m v_m}{\partial x} + \sum_i (-1)^i \alpha_i \frac{\partial}{\partial t}[(1-\alpha_i)\Gamma J] \quad \text{(SM-23)}$$

The second term of Eq. (SM-21) gives:

$$\sum_i \alpha_i \frac{\partial \Xi_i}{\partial x} = j\Delta\rho \frac{\partial J}{\partial x} - j\sum_i (-1)^i \alpha_i \frac{\partial}{\partial x}[(1-\alpha_i)\Gamma J] + \frac{1}{2}\sum_i \alpha_i \frac{\partial}{\partial x}(\Gamma_i J^2) \quad \text{(SM-24)}$$
$$+ \frac{\partial \varphi_\alpha}{\partial x}$$

Combining Eqs. (SM-21) to (SM-24):

$$\frac{\partial \rho_m v_m}{\partial t} + j\frac{\partial \rho_m v_m}{\partial t} + \sum_i (-1)^i \alpha_i \left(\frac{\partial}{\partial t} + j\frac{\partial}{\partial x}\right)[(1-\alpha_i)\Gamma J] - j\Delta\rho\frac{\partial J}{\partial x} \quad \text{(SM-25)}$$
$$- \frac{1}{2}\sum_i \alpha_i \frac{\partial}{\partial x}(\Gamma_i J^2) - \frac{\partial \varphi_\alpha}{\partial x} = 0$$

Using Eq. (7), the third term of Eq. (SM-25) reduces to:

$$\sum_i (-1)^i \alpha_i \left(\frac{\partial}{\partial t} + j\frac{\partial}{\partial x}\right)[(1-\alpha_i)\Gamma J]$$
$$= \left(\frac{\partial}{\partial t} + j\frac{\partial}{\partial x}\right)\left[\Gamma J \sum_i (-1)^i \alpha_i(1-\alpha_i)\right]$$
$$- \Gamma J \sum_i (-1)^i (1-\alpha_i)\left(\frac{\partial \alpha_i}{\partial t} + j\frac{\partial \alpha_i}{\partial x}\right) \quad \text{(SM-26)}$$
$$= \left(\frac{\partial}{\partial t} + j\frac{\partial}{\partial x}\right)\left[\Gamma J \sum_i (-1)^i \alpha_i(1-\alpha_i)\right] + \Gamma J \frac{\partial J}{\partial x}\sum_i (1-\alpha_i)$$
$$= \left(\frac{\partial}{\partial t} + j\frac{\partial}{\partial x}\right)\left[\Gamma J \alpha_1 \alpha_2 \sum_i (-1)^i\right] + \frac{1}{2}\Gamma \frac{\partial J^2}{\partial x} = \frac{1}{2}\frac{\partial \Gamma J^2}{\partial x} - \frac{1}{2}J^2 \frac{\partial \Gamma}{\partial x}$$
$$= \frac{1}{2}\frac{\partial \Gamma J^2}{\partial x} - \frac{1}{2}J^2 \sum_i \Gamma_i \frac{\partial \alpha_i}{\partial x}$$

Combining Eqs. (SM-25) and (SM-26):

$$\frac{\partial \rho_m v_m}{\partial t} + j\frac{\partial}{\partial x}(\rho_m v_m - \Delta\rho J) + \frac{1}{2}\frac{\partial \Gamma J^2}{\partial x} - \frac{1}{2}\sum_i \left[J^2 \Gamma_i \frac{\partial \alpha_i}{\partial x} + \alpha_i \frac{\partial}{\partial x}(\Gamma_i J^2)\right] - \frac{\partial \varphi_\alpha}{\partial x} \quad \text{(SM-27)}$$
$$= 0$$

The summation in the fourth term of Eq. (SM-27) reduces to:

$$\sum_i \left[ J^2 \Gamma_i \frac{\partial \alpha_i}{\partial x} + \alpha_i \frac{\partial}{\partial x} (\Gamma_i J^2) \right] = \frac{\partial}{\partial x} \left( J^2 \sum_i \alpha_i \Gamma_i \right) = \frac{\partial}{\partial x} (\Gamma_m J^2) \qquad \text{(SM-28)}$$

Combining Eqs. (1), (3), (SM-27) and (SM-28) yields:

$$\frac{\partial \rho_m v_m}{\partial t} + \frac{\partial \rho_m v_m^2}{\partial x} + \frac{\partial}{\partial x} \left\{ \left[ \frac{1}{2} (\Gamma - \Gamma_m) - \frac{\Delta \rho^2}{\rho_m} \right] J^2 \right\} = \frac{\partial \varphi_\alpha}{\partial x} \qquad \text{(SM-29)}$$

Using the expressions of $\Gamma$ and $\Gamma_m$, Eqs. (SM-13) and (SM-14), the bracket in the third term of the r.h.s. of Eq. (SM-29) can be written as:

$$\frac{1}{2} (\Gamma - \Gamma_m) - \frac{\Delta \rho^2}{\rho_m} = \sum_i \frac{\rho_i}{\alpha_i} - \frac{\Delta \rho^2}{\rho_m} - \frac{1}{2} \sum_i \rho_i \alpha_i \left[ \gamma_i - \frac{d(\alpha_i \gamma_i)}{d\alpha_i} \right]$$
$$= \frac{\rho_1 \rho_2}{\alpha_1 \alpha_2 \rho_m} - \frac{1}{2} \sum_i \rho_i \alpha_i^2 \frac{d\gamma_i}{d\alpha_i} \qquad \text{(SM-30)}$$

Combining Eqs. (SM-29) and (SM-30), finally yields:

$$\frac{\partial \rho_m v_m}{\partial t} + \frac{\partial}{\partial x} \left[ \rho_m v_m^2 + \frac{\rho_1 \rho_2}{\alpha_1 \alpha_2 \rho_m} J^2 + \frac{1}{2} J^2 \sum_i \rho_i \frac{d\gamma_i}{d(1/\alpha_i)} \right] = -\frac{\partial p}{\partial x} \qquad \text{(SM-31)}$$

Note that the pressure is recognized as the scalar field $p = -\varphi_\alpha$ in Eq. (SM-29). The emergence of the pressure field in this fashion is typical in variational formulations of incompressible problems.

**Equation of the relative motion between fluids**

From Eqs. (SM-19) and (SM-20), we verify that:

$$\sum_i (-1)^i U_i = \Gamma J - j\Delta\rho \qquad \text{(SM-32)}$$

$$\sum_i (-1)^i \Xi_i = \frac{1}{2} \Gamma' J^2 - \Gamma jJ \qquad \text{(SM-33)}$$

Hence, the transport equation of the relative motion between fluids is obtained by subtracting Eqs. (SM-18), which gives:

$$\frac{\partial \mathcal{M}_J}{\partial t} + \frac{\partial \Sigma_J}{\partial x} = 0 \qquad \text{(SM-34)}$$

where:

$$\mathcal{M}_J = \Gamma J - j\Delta\rho \qquad \text{(SM-35)}$$

$$\Sigma_J = -\frac{1}{2} \Gamma' J^2 + \Gamma jJ \qquad \text{(SM-36)}$$

**Conservative forces**

Conservative forces can be included by subtracting in the Lagrangian the potential energy from which these forces derive. For upward vertical flow, the gravitational potential is given by:

$$G = g\rho_m x \qquad \text{(SM-38)}$$

This introduces the following term in the derivatives of the Lagrangian:

$$\frac{\partial(-G)}{\partial \alpha_i} = -\rho_i g x \qquad \text{(SM-39)}$$

which adds a force $g\rho_m$ to the mixture momentum equation, and a force $g\Delta\rho$ to the equation of the relative motion.

**Interfacial drag**

Non-conservative interactions between fluids are modeled by means of a drag force $\mathfrak{D}_J$:

$$\mathfrak{D}_J = \frac{C_D}{D}\rho_1 u_r |u_r| \qquad \text{(SM-45)}$$

In order to keep the values of the variables within ranges typically encountered in real situations, we produce the interfacial drag coefficient $C_D$ from a set of experimental data of the relative velocity $u_r$ measured at different $\alpha_2$, available in the open literature [13, 14].

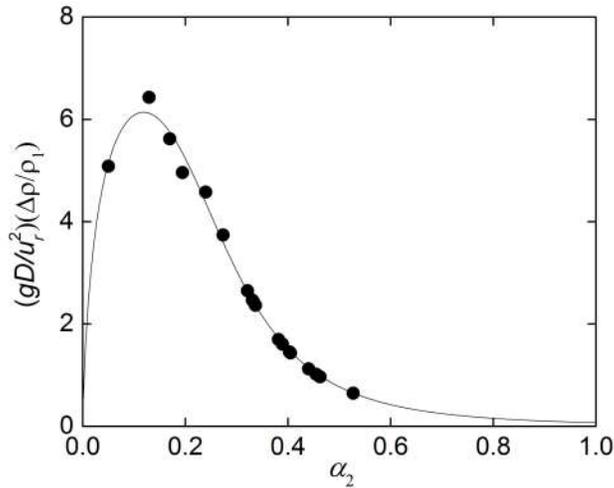

Figure 6. Drag coefficient constitutive law. (●) Experimental data [14,15], (solid curve) Eq. (SM-47).

At steady state, the drag and buoyancy forces balance each other, and therefore we can write:

$$C_D(\alpha_2) = \frac{gD\,\Delta\rho}{u_r^2(\alpha_2)\rho_1} \tag{SM-46}$$

Figure 6 shows the phenomenological function fitted together with the experimental data. The resulting drag coefficient function is:

$$C_D = \exp(1.32 + 8.4\alpha_2 - 42\,\alpha_2^2 + 45\alpha_2^3 - 15.4\alpha_2^4) \tag{SM-47}$$